\documentclass[conference]{IEEEtran}
\ifCLASSINFOpdf
   \usepackage[pdftex]{graphicx}
\else
   \usepackage[dvips]{graphicx}
\fi
  \usepackage[caption=false,font=footnotesize]{subfig}
\usepackage{url}
\graphicspath{{pdf/}{jpeg/}{png/}{eps/}{figures/}}
\DeclareGraphicsExtensions{.pdf,.jpeg,.png,.eps}
\usepackage{algorithmic}
\usepackage[table]{xcolor}
\usepackage{multirow}
\usepackage{listings}
\usepackage{stfloats}
\usepackage{balance}
\usepackage{soul}
\usepackage[normalem]{ulem}
\usepackage{cancel}

\begin{document}
\title{In Vivo Evaluation of the Secure Opportunistic Schemes Middleware using a Delay Tolerant \\ Social Network%A Case Study on the Need for an Opportunistic Routing testbed for \textit{In vivo} Evaluation%testbed for Delay Tolerant Social Networks
}

%\author{\IEEEauthorblockN{Michael Shell}
%\IEEEauthorblockA{School of Electrical and\\Computer Engineering\\
%Georgia Institute of Technology\\
%Atlanta, Georgia 30332--0250\\
%Email: http://www.michaelshell.org/contact.html}
%\and
%\IEEEauthorblockN{Homer Simpson}
%\IEEEauthorblockA{Twentieth Century Fox\\
%\and
%\IEEEauthorblockN{James Kirk\\ and Montgomery Scott}
%\IEEEauthorblockA{Starfleet Academy\\
%San Francisco, California 96678--2391\\

% \author{
% \IEEEauthorblockN{Corey E. Baker} 
% \IEEEauthorblockA{
% %Department of Electrical \\ 
% %and Computer Engineering, 
% University of California\\
% Email: cobaker@eng.ucsd.edu
% }
% \and
% \IEEEauthorblockN{Allen Starke} 
% \IEEEauthorblockA{
% %Department of Electrical \\
% %and Computer Engineering \\
% University of Florida\\
% Email: allen1.starke@ufl.edu
% }
% \and
% \IEEEauthorblockN{Shitong Xing}
% \IEEEauthorblockA{
% %Department of Electrical \\ 
% %and Computer Engineering, 
% University of California, San Diego\\
% Email: sxing@ucsd.edu
% }
% \and
% \IEEEauthorblockN{Janise McNair}
% \IEEEauthorblockA{
% %Department of Electrical \\
% %and Computer Engineering \\
% University of Florida\\
% Email: mcnair@ece.ufl.edu
% }}

\author{\IEEEauthorblockN{Corey E. Baker\textsuperscript{1}, Allen Starke\textsuperscript{2}, Tanisha G. Hill-Jarrett\textsuperscript{3}, Janise McNair\textsuperscript{2}}
\IEEEauthorblockA{
\textsuperscript{1}Department of Electrical and Computer Engineering, University of California, San Diego\\
\textsuperscript{2}Department of Electrical and Computer Engineering, University of Florida\\
\textsuperscript{3}Department of Clinical \& Health Psychology, University of Florida\\
Email: cobaker@eng.ucsd.edu, allen1.starke@ufl.edu, thilljarrett@phhp.ufl.edu, mcnair@ece.ufl.edu}}


%Telephone: (800) 555--1212, Fax: (888) 555--1212}


\maketitle

\begin{abstract}

Over the past decade, online social networks (OSNs) such as Twitter and Facebook have thrived and experienced rapid growth to over 1 billion users. 
%OSNs have become a major tool for 21st century message dissemination and offers a platform ranging from  advertising to  socializing with friends to social justice response.
%
%Unfortunately, as the name implies, OSNs require Internet availability to function, redu
A major evolution would be to leverage the characteristics of OSNs to evaluate the effectiveness of the many routing schemes developed by the research community in real-world scenarios. %opportunistically 
%in developing communities that have limited Internet access or during times when centralized infrastructures are compromised. %Recent products like WiFi Direct, Bluetooth, and LTE-D, allow mobile devices to communicate directly with each other without the need of an access point or cellular tower. However, with these devices, messages are local and do not travel out of the aforementioned radio communication ranges of the originating devices, e.g., from phone to printer, or for gaming in a co-located group. %D2D communication alone does not address the need of delivering information in an intermittent social group.
In this demonstration, we showcase the Secure Opportunistic Schemes (SOS) middleware which allows different routing schemes to be easily implemented relieving the burden of security and connection establishment. 
The feasibility of creating a delay tolerant social network is demonstrated by using SOS to enable AlleyOop Social, a secure delay tolerant networking research platform that serves as a real-life mobile social networking application for iOS devices. AlleyOop Social allows users to interact, publish messages, and discover others that share common interests in an intermittent network using Bluetooth, peer-to-peer WiFi, and infrastructure WiFi.  
\\

%To validate the capabilities of AlleyOop Social, a small-scale (10 devices) field experiment was conducted. 
%In addition, a case study was conducted to identify possible settings in which the AlleyOop Social app is desirable.
  
\end{abstract}

% \begin{IEEEkeywords}
% opportunistic communication, delay tolerant networking, routing, device-to-device, peer-to-peer, ad-hoc, Bluetooth, WiFi, social network, middleware, mobile applications, research platform, real-world evaluation, testbeds
% \end{IEEEkeywords}

% \fi
\IEEEpeerreviewmaketitle

\section{Introduction}\label{sec_intro}
Over the past decade, online social networks (OSNs) such as Twitter and Facebook 
have thrived and experienced rapid growth to over 1 billion users~\cite{faloutsos2010online}.
%
%Originally, OSNs were mainly used for establishing social connections with 
%friends, colleagues and family. 
%
%However, OSNs have grown in their application and are now used to access 
%large numbers of people for forming social groups, entertainment, natural 
%disasters, or responding to social justice issues. 
%
%A major evolution of these networks would be to disconnect them from the 
%Internet to create a true, people-centric and opportunistic social 
%network. 
%
A major limitation of OSNs is the dependence on Internet which is often 
sparse, difficult to maintain, or unavailable in rural areas or developing 
communities. 
In developed communities with cellular infrastructure, networks can become 
overwhelmed by too many users, particularly during emergencies.
In natural disaster situations, Internet and cellular 
communication infrastructures can be severely disrupted, prohibiting users from 
notifying family, friends, and associates about safety, location, food, water, and 
other resources. 
In addition, natural disasters typically damage infrastructure, which increases network traffic demands 
on any available undamaged infrastructure, causing congestion and delays.
%

%-----------------------------------------------------------------------------
Opportunistic communication can seamlessly supplement Internet connectivity 
when needed and keep communication channels open even during high-use and 
extreme situations.
%
% The proliferation of smart cities and the Internet of Things (IoT) encourage information 
% sharing between devices and sensors. 
% %
% Coupling smart cities and opportunistic communication could provide a low-cost 
% solution for delivering and relieving data demands of centralized infrastructures. 
%The dissemination of messages would therefore be possible during natural disasters, 
%when seeking emergency medical care in a crowded stadium where wireless 
%access is overloaded, or when issuing warnings of imminent danger to disconnected
%groups of people during hazardous or nefarious incidents.
%
%
%This would open a host of possibilities for the use of new real-time 
%applications in critical situations when no Internet or cellular service is 
%available. 
%
Furthermore, opportunistic communication can also serve as a low-cost solution for smart cities,
allowing developing and metropolitan areas to route smart city data through mobile 
and stationary nodes such as pedestrians, vehicles, street lights, public transportation.
DTN routing has the ability to deliver data in an intermittent network, 
but a major challenge for DTN routing is assessing real-world 
performance~\cite{hui2008phase,Keraenen2009,Baker2013}. 
To truly understand the reliability of DTNs and their ability to support social networks, it is imperative that DTN routing schemes are evaluated 
\textit{in vivo} with use-cases that are replicable, comparable, and available 
to a variety of researchers. 

In this demonstration, we present the Secure Opportunistic Schemes (SOS) middleware, a 
novel middleware that facilitates secure message delivery in cases where 
mobile connectivity is limited, unavailable, or non-existent. 
The SOS middleware supports real-life delay tolerant social networks on 
mobile devices. 
This allows mobile devices to leverage SOS to dynamically deliver messages to 
interested nodes when network infrastructure is not available and improve 
message delivery when infrastructure is available.
Additionally, the AlleyOop Social research platform is leveraged, 
which serves as an overlay application for SOS to create a delay tolerant 
social network for Apple iOS devices~\cite{baker2017demo}. 
%
% AlleyOop Social refers to the sports basketball play (alley oop) where one 
% player throws the ball close to the basket, but it is not able to reach 
% the final destination. While the ball is in flight, a player that is closer 
% to the basket catches the ball and then is able to deliver the ball to its 
% final destination. 
% %
% In the same regard, AlleyOop Social enables wireless mobile users to 
% communicate over longer distances by sending messages that cannot reach 
% the final destination, but are ``caught'' by intermediate mobile devices, 
% which continue to catch and pass the messages until they are delivered to 
% the final destination.
%
AlleyOop Social is named after the basketball play known as an ``alley oop''. 
An ``alley oop'' occurs when one player throws the ball close to the basket, but it 
is not able to reach the final destination. 
While the ball is in flight, a teammate that is closer 
to the basket catches the ball and scores. 
In the same regard, AlleyOop Social enables wireless mobile users to 
communicate over longer distances by sending messages that cannot reach 
the final destination, but are ``caught'' by intermediate mobile devices, 
which continue to catch and pass the messages until they are delivered to 
the final destination.

\begin{figure}
	\centering
	\includegraphics[width=.48\textwidth, height=2.5in]{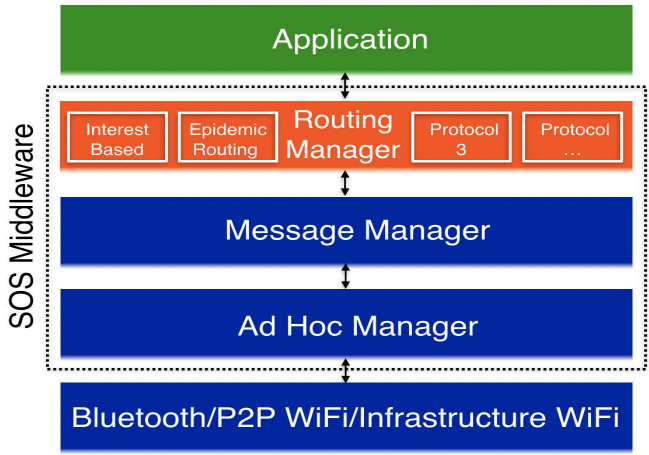}
	\caption{SOS middleware system stack - green represents mobile applications created by developers, orange represents the modular routing layer consisting of multiple opportunistic schemes created by academic researchers, blue represents foundational layers in the middleware that consist of encryption, authentication, peer discovery, connection management, and data dissemination. Managers marked with the color ``blue'' cannot be modified by mobile application developers or academic researchers.}
	\label{fig_alleyoop_sys_stack}
\end{figure}

\section{Related Work}\label{related_work}
%-----------------------------------------------------------------------------
%Various middlewares~\cite{su2007haggle, Juste2013a, dubois2015supporting}, testbeds~\cite{doering2008ibr,paul2015challenges,moraleda2014link}, and mobile applications have been developed to 
%address the issues of providing deployable delay tolerant networking 
%applications which can operate with minimal infrastructure and effectively 
%evaluate DTN routing protocols.
% various middlewares, testbeds, and 
%mobile applications have been developed.
%
%The rest of Section~\ref{related_work} will highlight some of 
%the related contributions to deployable delay tolerant and ad hoc applications 
%and testbeds.
In recent years, a number of social-aware and social-based routing schemes have
leveraged social interactions to deliver data using delay tolerant networks 
(DTNs)~\cite{wei2014survey}. 
However, related work has primarily evaluated routing protocols in simulation 
environments, which provide valuable analyses, but are based on synthetic 
mobility patterns to emulate node movement and tend to use abstract models to imitate the radio response of real commodity wireless technologies\cite{hui2008phase,Keraenen2009,Baker2013}.
There are a few studies that have taken on the approach of demonstrating DTNs 
in realistic environments~\cite{doria2009providing,pentland2004daknet,doering2008ibr}. 
However, these studies do not consider other significant aspects, such as user 
security and privacy along with the limitation of operating with only the 
epidemic routing scheme. 
Various middlewares~\cite{pietilainen2009mobiclique,helgason2010mobile,skjegstad2012mist,
dubois2015supporting}, testbeds~\cite{mukhtar2006backup,morgenroth2012bundle,juste2014tincan, moraleda2014link, ben2014demo, paul2015challenges, siby2015method, asadpour2016route}, 
and mobile applications have been developed to 
address providing deployable delay tolerant networking applications which can 
operate with minimal infrastructure and effectively evaluate DTN routing 
protocols. %The purpose of this sections is to provide a brief overview on 
%middlewares and testbeds developed by researchers.

\section{Secure Opportunistic Schemes (SOS) Middleware}\label{sec_sos_middleware}
%-----------------------------------------------------------------------------
The SOS middleware is an underlying framework that turns the AlleyOop Social 
research platform into a delay tolerant mobile social network.
The SOS middleware takes a modular approach to abstract away much of the 
complexity involved in implementing opportunistic routing schemes such as 
device dicovery, establishing D2D connections, and handling device security and 
privacy.
%
%The MPC framework in Section~\ref{alleyoop_subsec_ios} is incorporated into the
%{Ad Hoc Manager} which directly oversees the transport, network, data link, 
%and physical layers.%, AlleyOop will be primarily concerned with happenings above the transport layer.
%
DTNs are intended to provide an overlay architecture above the existing 
transport layer and ensure reliable routing during intermittency~\cite{Fall2003}.
Building on the knowledge gained from previous middlewares~\cite{su2007haggle,caporuscio2012ubisoap}, SOS hides the complexity of the network stack (session and 
presentation) within the ad hoc manager, message manager, and routing manager,
allowing any mobile application to run at the application layer as depicted in 
Figure~\ref{fig_alleyoop_sys_stack}.
Different from other middlewares such as the Haggle Project~\cite{su2007haggle},
a separate instance of the SOS middleware is intended to run within each mobile application 
as opposed to a daemon which often requires devices to be rooted or jailbroken.
Designing SOS in this manner allows for the middleware to be integrated within any 
mobile application in iOS, enabling them to support opportunistic communication 
without jailbreaking devices along with being compliant with App Store regulations.

\subsection{Application}\label{alleyoop_subsubsec_alleyoop_application}
%-----------------------------------------------------------------------------
Mobile applications serve as an overlay to the SOS middleware. 
Applications can be of any form such as social networking, medical, or any 
other type of application that would like to share data opportunistically.
%
%To provide a user interface that will attract interest and inspire usage, the 
%{AlleyOop Social Application} layer is designed using storyboards in Xcode.
%
Mobile applications are responsible for providing a user interface 
to users and storing data to local or online storage systems.  
The SOS Middleware provides a number of API's for sending/receiving data, 
surrounding user notification, routing protocol selection, and security and 
privacy preferences.
Existing mobile applications can simply add the SOS middleware as a framework 
and start using the aforementioned API's to send and receive data.
Applications are responsible providing the data to be sent as well as 
handling data once it has been received and decrypted.

\subsection{Routing manager}\label{alleyoop_subsubsec_routing}
%-----------------------------------------------------------------------------
The {routing manager} is responsible for leveraging D2D connections to 
transform any application into a delay tolerant networking application that 
delivers messages to out of range nodes in the midst of intermittency.
Routing in SOS is designed for modularity, permitting additional 
DTN routing schemes to be developed on top of the {message manager} and run 
seamlessly under the {Application} layer.
Designing SOS in this manner allows for a flexible middleware that enables 
applications to dynamically change based on user preference without the 
need of modifying hardware or other layers in the software stack.
Currently, the {routing manager} in SOS has two DTN routing protocols 
implemented: epidemic routing and interest-based routing.
Epidemic routing is a simple routing scheme that achieves effectiveness through 
gratuitous replication and delivery of messages upon node encounters~\cite{Vahdat2000}.
The IB routing protocol %is available in AlleyOop Social which 
operates in a similar manner to epidemic routing, except, instead of 
propagating messages to all users, messages are only propagated to interested 
users who are subscribed to the publisher of the original message.
Due to the modular nature of the SOS middleware, additional routing 
protocols can be added to the {routing manager}.
APIs are available to all protocols in the {routing manager} to 
facilitate communication between the {message manager} and the 
{application} layer.
Both the IB and Epidemic routing protocols are written in less than 100 lines 
of Swift code. %and can be found on GitHub\footnote{https://github.com/}. 

\subsection{Message manager}\label{alleyoop_subsubsec_message_manager}
%-----------------------------------------------------------------------------
The {message manager} notifies the respective protocol used in the 
{routing manager} whenever a new peer has been discovered or lost. 
Additionally, the {message manager} is responsible for taking action whenever a 
connection state changes. 
For example, if the disconnection between two users is lost, the {message manager}
knows what messages were not transferred.
Lastly, the {message manager} translates messages between the {routing manager} 
and {ad hoc manager} in a common format for both layers to interpret.

\subsection{Ad hoc manager}\label{alleyoop_subsubsec_adhoc_manager}
%-----------------------------------------------------------------------------
The {ad hoc manager} manages Apple's multipeer connectivity (MPC) framework, 
which allows communication between iOS, macOS, and tvOS devices using 
peer-to-peer WiFi, Bluetooth personal area networks, or infrastructure WiFi networks\footnote{Apple Inc., Multipeer connectivity framework reference, https://developer.apple.com/library/ios/documentation/MultipeerConnectivity/ Reference/MultipeerConnectivityFramework/}.
To the best of our knowledge, SOS is the first middleware to leverage MPC to evaluate multiple delay tolerant routing schemes.
The {ad hoc manager} is responsible for viewing discovered peers, establishing 
D2D connections, encrypting connections, encrypting data from end-to-end, 
generating keys, validating certificates, as well as signing and verifying 
data sent and received data.
%
%The {Ad Hoc Manager} also handles key generation and retrieval,  
%generation of certificate signing requests (CSRs), and 
%storage/retrieval of certificates.
%
Apple's documentation on how to use MPC is detailed, but the company does 
not disclose specific details on how MPC works.
For example, specifics about the encryption methods MPC uses are not provided.
Details about how the SOS middleware handles security and privacy are elaborated 
on in Section~\ref{alleyoop_sec_privacy}.  

\begin{figure*}%[htbp]
\centering
%Enabling offline security: one-time infrastructure requirement
    \subfloat{\includegraphics[width=0.48\textwidth, height=2.7in]{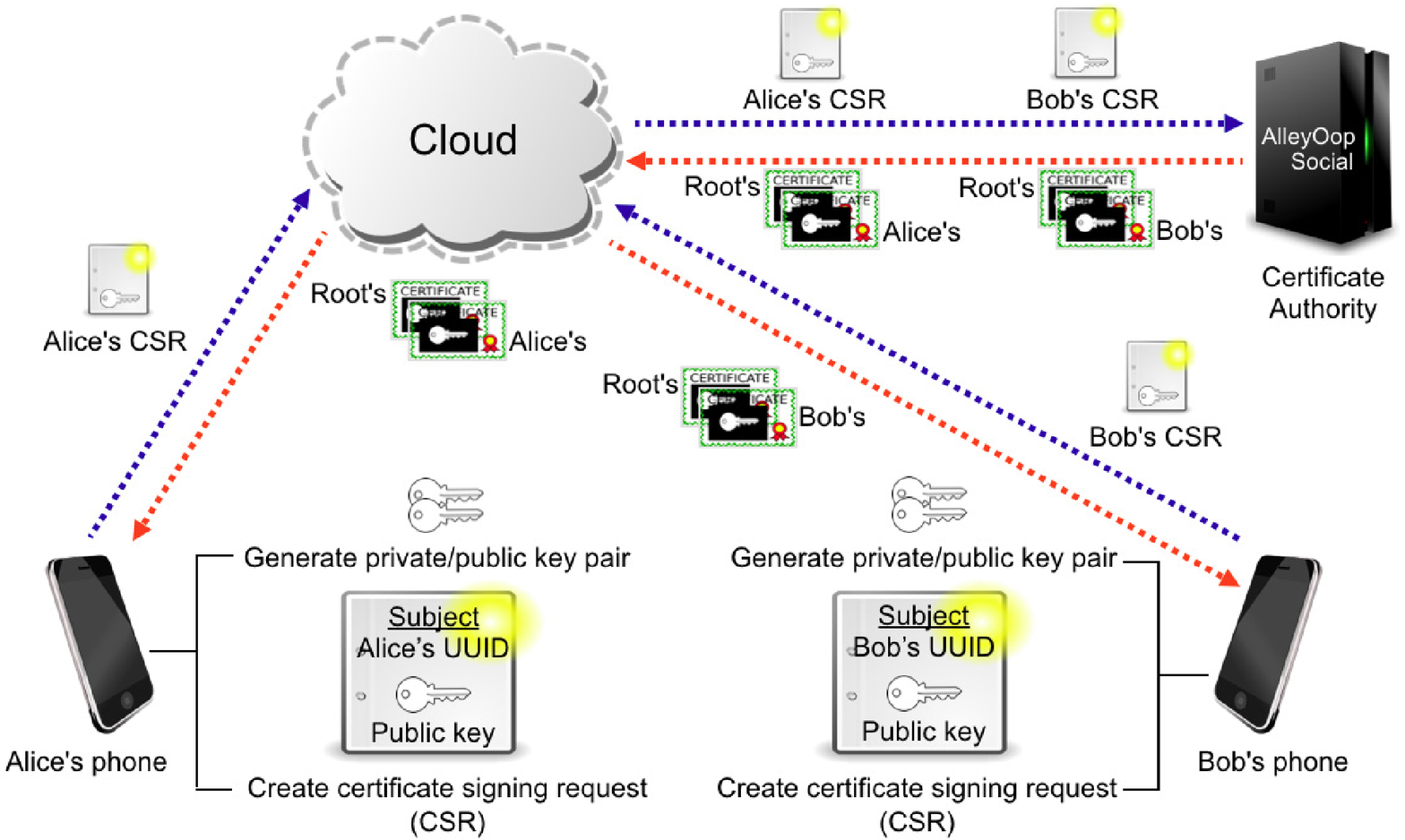}\label{fig_alleyoop_sign_up}} 
    ~%\hfill
	\subfloat{\includegraphics[width=0.48\textwidth, height=2.7in]{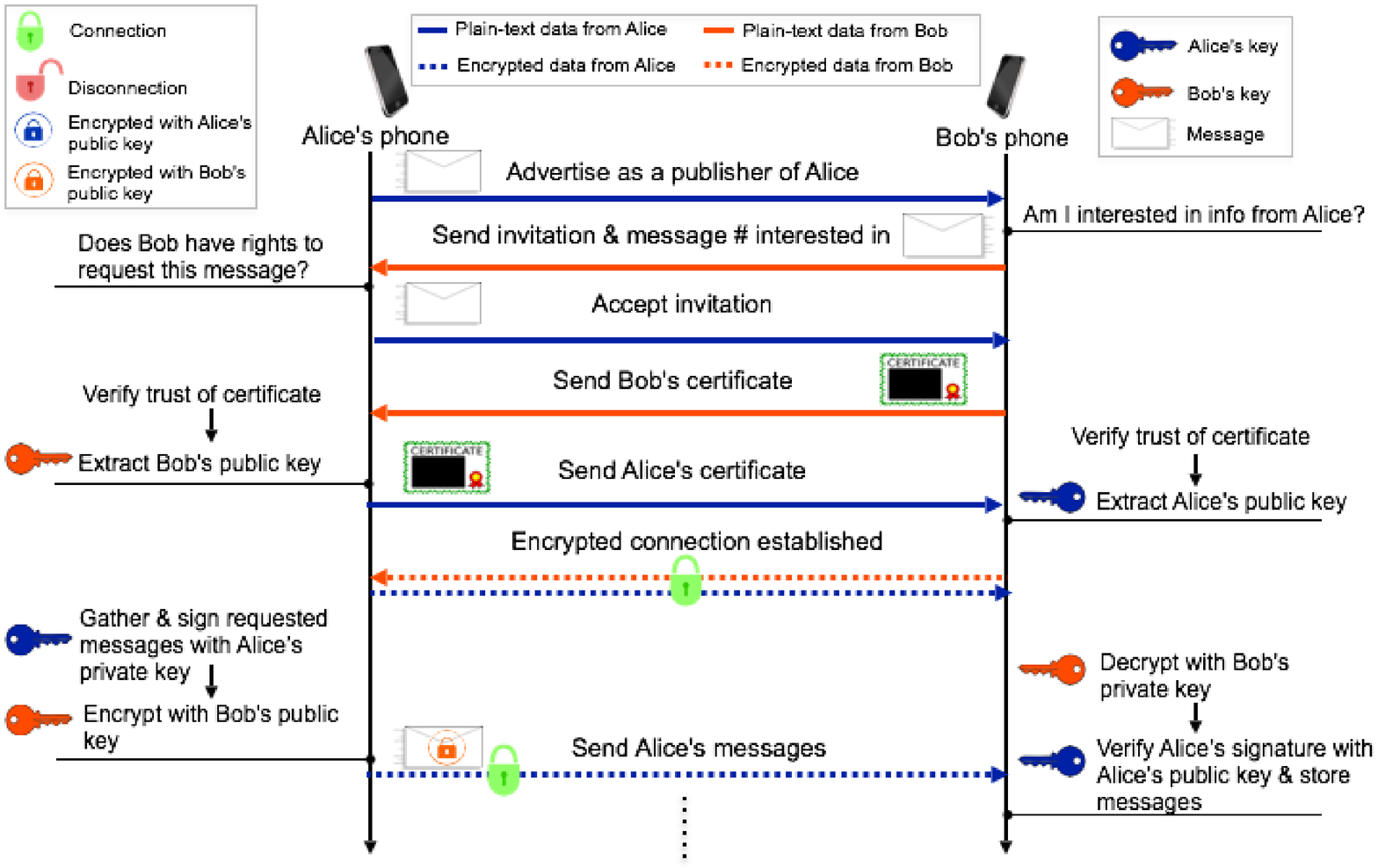}\label{fig_allleyoop_decentralized}} 
    %Decentralized communication
   
\caption{(a) One-time infrastructure requirement, occurs during account creation to enable DTN security. (b) Decentralized communication between nodes.} %AlleyOop Social secure message dissemination
\label{fig_alley_security}
\end{figure*}

\section{Privacy and Security}\label{alleyoop_sec_privacy}
%-----------------------------------------------------------------------------
In regard to network security there is no ``one-size-fits-all" 
approach~\cite{perlman1999overview}.
Security concerns may become exacerbated in delay tolerant and ad hoc 
applications where nodes are vulnerable to attacks such as eavesdropping, 
denial of service, and compromised devices.
Providing secure communication that prevents an adversary from accessing and/or 
modifying data is a fundamental requirement of any DTN application~\cite{cabaniss2015multi}.
%
% The researchers who designed the Haggle project~\cite{su2007haggle} but explain
% that security is a proof-of-concept and only discuss how security can be 
% implemented, and makes no claim their implementation is secure.
Previous research discusses security in opportunistic applications conceptually 
and makes no claims that the implementations are secure~\cite{su2007haggle}.
The intent of the section is to provide a novel, but simple concept and 
implementation of an initial layer of security for DTN protocols and enable 
the overlaying mobile application to detect the identity of its users, send encrypted 
information, verify the originating source of the information being forwarded, and ensure that data have not been modified --- all with minimal dependence on centralized infrastructures.

%-----------------------------------------------------------------------------
Additional security can be added to AlleyOop Social by incorporating mechanisms 
such as distributing CA functionality amongst nodes~\cite{kong2001providing}, 
or integrating trust measurements within the routing schemes~\cite{kumar2010protect} 
available in the routing manager discussed in Section~\ref{sec_sos_middleware}. 
To enable the initial layer of security in the SOS middleware, AlleyOop Social 
leverages conventional public-key infrastructure (PKI) techniques to create a one-time 
PKI requirement that occurs during initial download and user-signup for the application.
AlleyOop Social assumes that users will have Internet connectivity 
during the initial download and installation of the mobile app.
After the one-time infrastructure requirement, Internet connectivity is 
no longer needed for privacy, security, and message dissemination. 
The process of generating keys and receiving X.509 certificates in AlleyOop Social's 
one-time infrastructure requirement is depicted in 
Figure~\ref{fig_alleyoop_sign_up}. 

Using the one-time infrastructure requirement in Figure~\ref{fig_alleyoop_sign_up} is not without limitations.
The obvious shortfall is the ``one-time'' requirement.
A fair assumption is that the AlleyOop Social application along with others 
using the SOS middleware will acquire their mobile applications from the Apple 
App Store, which currently requires an Internet connection.
Assuming users sign up shortly after acquiring the application addresses some 
concerns with the ``one-time'' requirement.
Additionally, if a connection between a device and the cloud is somehow 
compromised, or a malicious device attempts to provide someone else's 
unique user-identifier during user sign-up, a certificate with the wrong 
credentials could be generated by the CA.
To circumvent this issue, the cloud can ask the CA to compare 
and validate the unique user-identifier provided in the certificate with the 
unique user-identifier affiliated with the logged in user.
Other limitations are also prevalent with the current security scheme such as 
an Internet connection is required to revoke specific user certificates, update 
CA root certificates, replenish expired certificates, and notify users of known 
malicious devices. 

\begin{figure*}[htbp]
 	\centering
 	\subfloat[AlleyOop Social message forwarder selection]{\includegraphics[width=0.45\textwidth, height=2.7in]{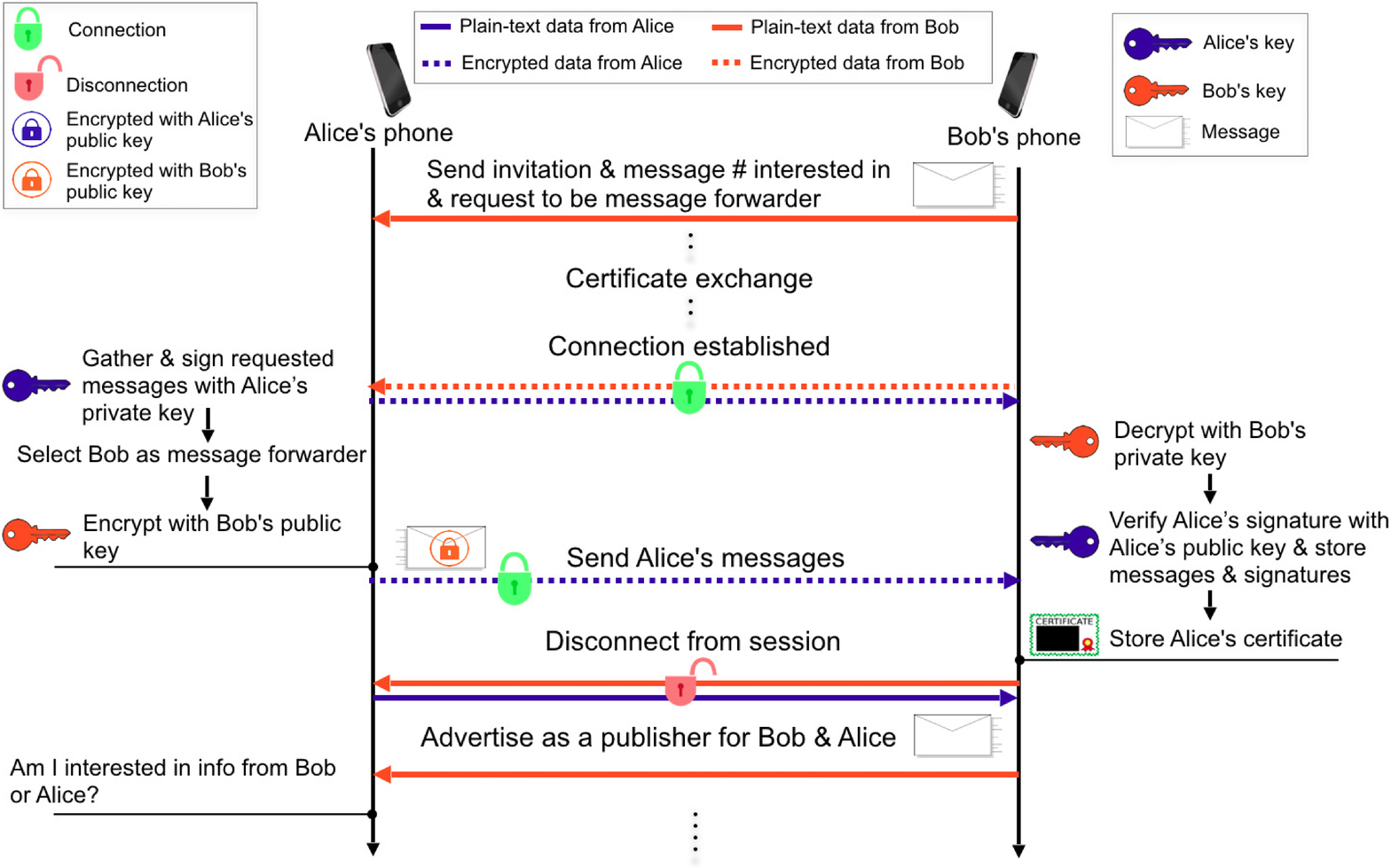}
 	%\caption{AlleyOop Social message forwarder selection}
 	\label{fig_alleyoop_msg_forwarder_selection}}
     ~
     \subfloat[AlleyOop Social message forwarder dissemination]{\includegraphics[width=0.45\textwidth, height=2.7in]{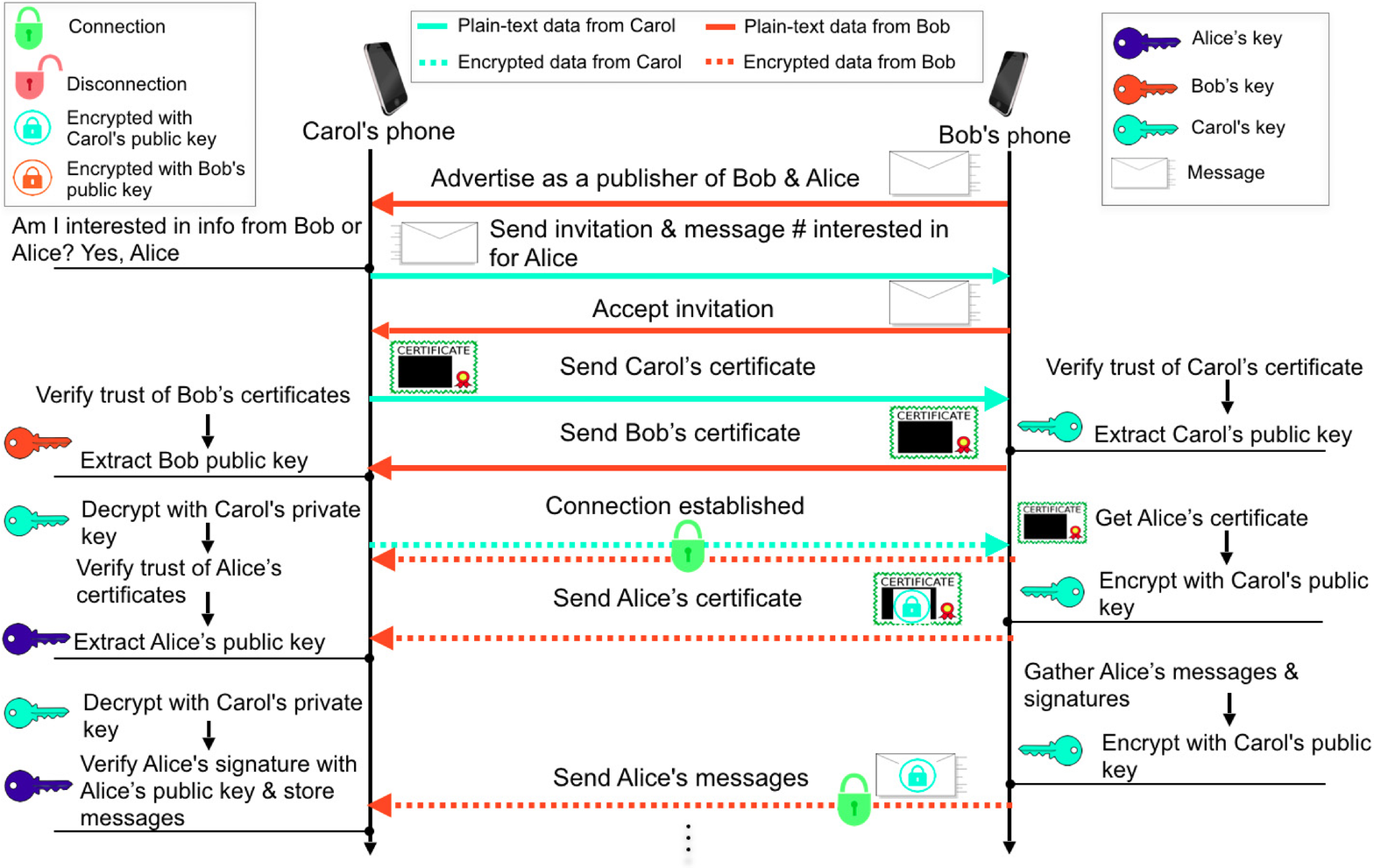}
 	%\caption{AlleyOop Social message forwarder dissemination}
 	\label{fig_alleyoop_msg_forwarder_diss}}    
    
     \caption{AlleyOop Social message forwarder}
 	\label{fig_alleyoop_msg_forwarder}    
    
\end{figure*}

\section{Message Dissemination}\label{alleyoop_sec_dissemination}
%-----------------------------------------------------------------------------
After sign-up is complete and a mobile device receives its respective certificate and 
AlleyOop Social CA root certificate, the user can disseminate messages to other AlleyOop Social users
using any DTN routing protocol discussed in Section~\ref{alleyoop_subsubsec_routing}.
Whenever a user creates a message or performs an action such as follow/unfollow 
of a user, AlleyOop Social performs the following two operations: 1) saves the action to the local database 
on the mobile device and 2) synchronizes the action with the cloud when the Internet 
becomes available.
Once an action is saved to the local database of the device it can 
be disseminated using a DTN routing protocol to interested AlleyOop Social users without the use of Internet. 
The following sections expound upon how messages are disseminated
after being created by a user in the {AlleyOop Social application} layer and passed 
to the {routing manager}.

\subsection{Advertisements and node discovery}\label{alleyoop_subsec_advertise}
%-----------------------------------------------------------------------------
Mobile devices roam freely advertising and browsing for basic information in plain-text to 
assist other AlleyOop Social enabled devices with making the decision of whether or not to request 
a connection.  
For example, the epidemic and IB routing protocols discussed in Section~\ref{alleyoop_subsubsec_routing} 
advertises a plain-text key/value dictionary consisting of \textit{UserID/MessageNumber}. 
%a: {[YyCVxDcerf: 8, ...:...]}.
%
The key field in the dictionary is a 10 byte unique user identification 
string.
The value field of the dictionary is the latest \textit{MessageNumber} that 
the advertising device has for the particular \textit{UserID}.
A browsing node is now able to quickly decide whether it is interested in
the \textit{MessageNumber} for the respective \textit{UserID} string and whether it 
should request a connection from the advertising node.
Figure~\ref{fig_allleyoop_decentralized} depicts a typical scenario in AlleyOop Social  
where Bob's device is interested in messages from Alice's device.

\subsection{Forwarder Selection \& Dissemination}\label{alleyoop_subsec_forwarder}
%-----------------------------------------------------------------------------
Depending on the DTN routing protocol being used in the {Routing} layer
of the  mobile device, prospective nodes can become message forwarders
for other users.
For example, in epidemic and IB routing, a node becomes a message 
forwarder for a particular user-identifier whenever a new message is requested 
and received.
When a node becomes a message forwarder, it follows a similar process to the 
one outlined in Figure~\ref{fig_allleyoop_decentralized}, with particular 
differences that are shown in Figures~\ref{fig_alleyoop_msg_forwarder_selection}.
% 
% For each message Bob receives from Alice, Bob stores the decrypted plain-text 
% message along with the signature of the message.
% %
% After Bob successfully receives all of Alice's messages, his device stores Alice's 
% certificate to his iOS keychain for future retrieval.
% %
% Bob then requests a disconnection and both devices return to advertising in plain-text.
%
Figure~\ref{fig_alleyoop_msg_forwarder_diss} shows the interaction between 
Bob and Carol when Carol is interested in Alice's message that Bob is forwarding.
The process is similar to message dissemination in 
Figure~\ref{fig_allleyoop_decentralized}, except Bob sends 
his certificate to Carol to establish an encrypted connection and in addition,
forwards Alice's certificate.

\begin{figure*}%[htbp]
\centering

    \subfloat[Social relationship directed graph for the ten (10) active users in Gainesville, FL]{\includegraphics[width=0.48\textwidth, height=2.2in]{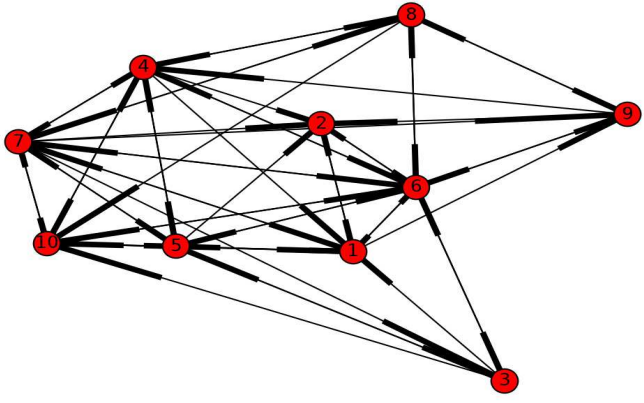}\label{fig_network_graph}} 
    ~
	\subfloat[AlleyOop Social map of ``active user'' message generation (blue) and message dissemination (red) in a $\sim$11km x 8km area]{\includegraphics[width=0.48\textwidth, height=2.2in]{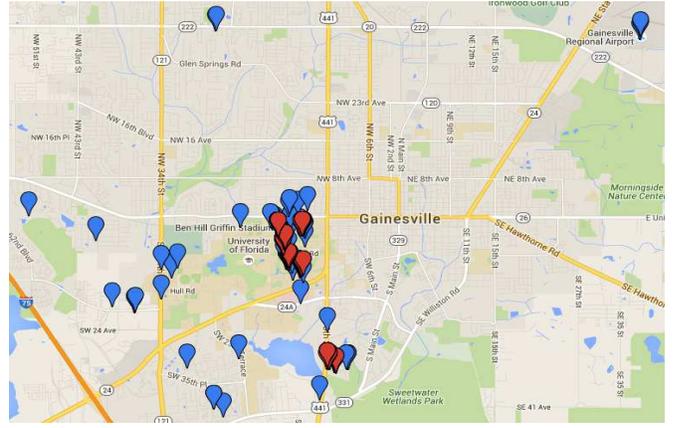}\label{fig_app_map}}

    \subfloat[Delay]{\includegraphics[width=0.48\textwidth, height=2.2in]{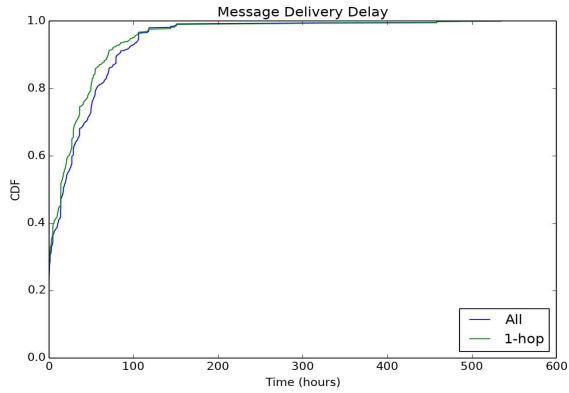}\label{fig_alley_delay}} 
    ~
    \subfloat[Delivery]{\includegraphics[width=0.48\textwidth, height=2.2in]{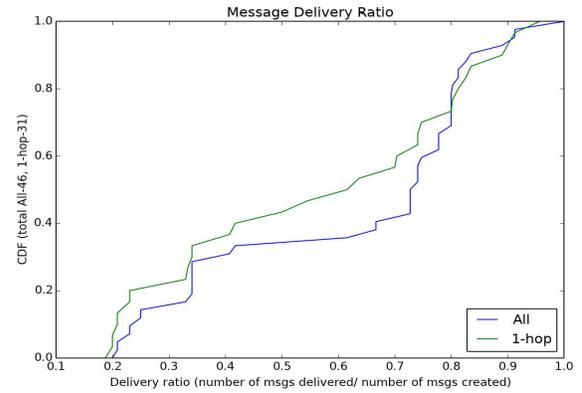}\label{fig_alley_deliv}}
    %~
    %\subfloat[AlleyOop Social map of message creation (green) and message dissemination (red)\cb{Need to fix. Only includes message dissemination points right now}]{\includegraphics[width=0.30\textwidth, height=2.0in]{app_map}\label{fig_app_map}}

\caption{AlleyOop Social real world evaluation results}
\label{fig_alley_results}
\end{figure*}

\section{Real World Evaluation}\label{alleyoop_sec_results}
%-----------------------------------------------------------------------------
%\input{comm_mag_results}
The AlleyOop Social application was available for beta testing in Apple TestFlight
app for $7$ days.
AlleyOop Social had thirty one ($31$) testers who downloaded the application around 
the United States.
%
%For users who opted-in, the locations of where messages were created and 
%disseminated can been seen in Figure~\ref{fig_app_map_all}.  
%
Due to limited amount of users and a critical mass of users who are socially 
related to each other, this section will constrain the results to users who 
passed at least one ($1$) D2D message using the IB routing protocol in Gainesville, FL. 
%
%Nine ($9$) of the users were undergraduate or graduate students at the University of Florida. %University 
%of Florida. 
%
%One of these student users had access to two devices, which resulted in ten 
%($10$) active devices in Gainesville, FL.
%
Ten devices used AlleyOop Social in a $\sim$11km x 8km area depicted in 
Figure~\ref{fig_app_map}, resulting in users posting $259$ unique messages.

\subsection{Social relationships}\label{subsec_social_relation}
%-----------------------------------------------------------------------------
Many of the students were friends before the field study and typically interacted 
during the school week.
% 
%AlleyOop Social users posted over $400$ messages ($310$ from active users) and created over $1200$ interactions (message passing and user subscriptions).
%
Individual users were given the freedom to choose other users to subscribe to;
therefore, all users did not follow each other. 
The digraph $\mathcal{G}(V,E)$ formed by the total nodes who 
participated $n = |V(\mathcal{G})| = 10$ is depicted in Figure~\ref{fig_network_graph}. 
A social relationship between a node pair $i,j~\in E$ is an edge
$e_{i,j}$, meaning that user $i$ follows user $j$.
The edge $e_{i,j}$ does not necessarily mean the edge $e_{j,i}$ exists because
some users did not follow each other back as in the case for node $1$ and node $3$
in Figure~\ref{fig_network_graph}.
The density of the social relationships is $0.64$, meaning that the majority 
of the possible social relationships were formed naturally by the participating 
nodes.
The compactness of $\mathcal{G}$ can be determined by calculating the average 
shortest path length between all node relationship pairs 
$\sum_{i\geq j}l(i,j) / \frac{n(n-1)}{2} = 1.3$, along with the maximum shortest 
path length, otherwise known as the diameter $d$ between any two nodes 
$d(\mathcal{G}) = \textrm{max}_{i,j\in V}~l(i,j) = 2$.
The compactness of the social relationship graph reveals that even if a user 
does not follow another user directly, there is still an indirect follower that is two
degrees away. 
The center nodes (6 and 7) of the social relationship graph has a radius of 
$1$ which reflects the nodes with the smallest eccentricity$_{i,j} = 
\textrm{max}_{i,j \in V}~l(i,j)$.

%-----------------------------------------------------------------------------
Additional features can be determined by translating Figure~\ref{fig_network_graph}
to a undirected graph.
This means that if a two-way relationship did not already exist, it will exist 
in the undirectional graph making $e_{i,j} = e_{j,i}$ for all $i,j\in E$.
Now the network transitivity is computed to be $T(\mathcal{G}) = 3*\textrm{ number of 
triangles }/\textrm{ number of connected triads} =  0.80$ which measures the extent
that a friend $k$ of a friend $j$ is also a friend of $i$.

\subsection{Message dissemination}\label{subsec_results}
%-----------------------------------------------------------------------------
The social relationship graph in Figure~\ref{fig_network_graph} provides an 
overall understanding of nodes' interests in messages along with providing insight 
into how nodes may cluster due to who they follow.
Figure~\ref{fig_network_graph} does not provide any insight on physical node 
locations or mobility during the evaluation.
Figure~\ref{fig_app_map} assists with understanding node mobility by showing 
where users created messages (blue) and passed messages (red) in Gainesville, FL.  
A total of $967$ messages were disseminated from user-to-user using IB routing 
in AlleyOop Social.
The total amount of subscriptions made by the ten (10) active users was $46$. 
Figure~\ref{fig_alley_delay} provide the delay results for 
messages disseminated via ``1-hop'' and ``All'' hops. 
%
%The delay in message delivery as well as the delivery ratio for are shown in Figures~\ref{fig_alley_delay} and~\ref{fig_alley_deliv} respectively. 
%
In regard to ``All'' messages, Figure~\ref{fig_alley_delay} shows that 
$0.43$ of the messages delivered had a delay of $24$ hours or less, 
while $0.90$ of the messages had a delay of 94 hours or less. 
In regard to ``1-hop'' delay, that $0.44$ of the messages delivered had a 
delay of $24$ hours or less, while $0.92$ of the messages had a delay of 94 
hours or less for ``1-hop'' messages. 
%
%We attribute the high delay to the lack of density in the evaluation\cb{calculate 2d density}.
%
%Coupling the knowledge of the node density with the knowledge of users going 
%to sleep, the delays in message delivery are reasonable for a real world 
%evaluation.  
%
%The delay and delivery ratios are only determined for users who came in direct contact with another AlleyOop Social user and disseminated at least one message using IB routing.

% \begin{figure}%[htbp]
% 	\centering
% 	\includegraphics[width=.45\textwidth, height=1.75in]{network_graph}
% 	\caption{Social connectivity graph of active users in Gainesville, FL. The center nodes are 2 and 7. The network has a density of 0.64 with the average shortest path being at 1.4}
% 	\label{fig_network_graph}
% \end{figure}

% \begin{figure}[htbp]
% 	\centering
% 	\includegraphics[width=.40\textwidth, height=1.75in]{app_map_all}
% 	\caption{AlleyOop Social map of ``all user'' message creation (blue) and message dissemination (red)}
% 	\label{fig_app_map_all}
% \end{figure}

%-----------------------------------------------------------------------------
In regard to message delivery, Figure~\ref{fig_alley_deliv} shows that $0.30$ 
of the subscriptions had a delivery ratio greater than $0.80$ for ``All'' messages.
$0.50$ of the subscriptions had a delivery ratio greater than $0.70$ for all 
messages.
$0.25$ of the subscriptions had a delivery ratio of $0.80$ for ``1-hop'' messages.  
Users delivered $0.826$ of the $967$ messages via 1-hop.
The additional $0.174$ were delivered using 2-hops or more and is depicted in ``All''.
%
%Nine out of $10$ active users delivered a portion of their own messages within 
%``1-hop''.
%
The compactness of the social relationships between the nodes 
discussed in Section~\ref{subsec_social_relation} partially explains why 
the majority of the messages were delivered within ``1-hop''.
Note the low density due to real people being able to operate freely in 
a large city area (88km$^{2}$), which resulted $0.93$ of the messages being delivered within in $94$ hours of creation.
DTN simulations typically model 50 to 100 nodes in a constrained simulation 
space ranging  between 0.25km$^{2}$ - 4km$^{2}$.
In addition, node mobility tends to become stationary, for at least 5-8 
hours a day due to the human requirement to sleep, thus limiting possible 
interactions between nodes.
The results at such a low density provide promising insight into delay tolerant social networks and suggest further investigations at higher densities are needed.

\section{Demonstration}\label{alleyoop_sec_demo}
%-----------------------------------------------------------------------------
During the demonstration attendees will be able to download AlleyOop 
Social on their iOS devices via Apple TestFlight.
Users can follow friends, post new messages, as well as toggle between DTN
routing schemes inside the application.
We will demonstrate both the online and offline modes by disconnecting mobile 
devices from cellular and WiFi networks.
\bibliographystyle{IEEEtran}
\bibliography{dtnreferences}

\end{document}